# A Connection Between Submillimeter Continuum Flux and Separation in Young Binaries

Eric L.N. Jensen,[1] Robert D. Mathieu,[1] Gary A. Fuller[2]

## ABSTRACT

We have made sensitive 800-$\mu$m continuum observations of low-mass, pre-main sequence (PMS) binary stars with projected separations less than 25 AU in Taurus-Auriga to study disks in the young binary environment. We did not detect any of the observed binaries, with typical $3\sigma$ upper limits of $\sim$ 30 mJy. Combining our observations with previous 1300-$\mu$m observations of PMS Taurus binaries by Beckwith et al. (1990) and others, we find that the submillimeter fluxes from binaries with projected separations $a_p$ between 1 AU and 50 AU are significantly lower than fluxes from binaries with $a_p > 50$ AU. The submillimeter fluxes from the wider binaries are consistent with those of PMS single stars. This may indicate lower disk surface densities and masses in the close binaries. Alternatively, dynamical clearing of gaps by close binaries is marginally sufficient to lower their submillimeter fluxes to the observed levels, even without reduction of surface densities elsewhere in the disks.

*Subject headings:* binaries: general — circumstellar matter — stars: formation — stars: pre–main sequence

## 1. Introduction

It is now well established that most pre-main sequence (PMS) and main sequence stars are members of binary systems (for reviews see Mathieu 1994 and Abt 1983). Massive disks likely play a central role in binary formation, for example as conduits transferring infalling material onto stars (e.g. Bonnell et al. 1992), or more fundamentally through production of companions in global disk instabilities (e.g. Adams, Ruden, & Shu 1989) or enhancement of capture cross-sections (Larson 1990).

[1]Univ. of Wisconsin-Madison, Dept. of Astronomy, 475 N. Charter St., Madison, WI 53706. E-mail: jensen@madraf.astro.wisc.edu, mathieu@madraf.astro.wisc.edu

[2]NRAO, 949 N. Cherry Ave., Campus Building 65, Tucson, AZ 85721. E-mail: gfuller@nrao.edu. Current address: NRAO, Edgemont Road, Charlottesville, VA 22903

Beckwith et al. (1990; BSCG) measured 1300-$\mu$m continuum emission from 86 PMS stars in Taurus-Auriga to search for disks. They, and later Beckwith & Sargent (1993), noted that in their sample massive disks (typically $\sim 0.02$ $M_\odot$) are rarely found in binary systems with projected separations $<$ 100 AU but are common in systems with wider projected separations. However, many of their upper limits on the disk masses of the closer binaries were comparable to the detected disk masses among the wider binaries. Also, numerous "single" stars in their sample have since been found to be binaries. In this Letter, we reevaluate this important finding of BSCG using a larger sample of binaries and more sensitive observations of the shortest-period systems.

## 2. Observations

Observations of 800-$\mu$m continuum emission were carried out using the James Clerk Maxwell Telescope (JCMT)[3] and UKT14, the facility bolometer. The observations were made under stable conditions on 1993 January 23 and 24, using a 65-mm aperture which corresponds to a beam size of 18″. Chopping and nodding were used to remove background emission, using a 7.8 Hz, 60″ chop throw in azimuth. Pointing and focus of the telescope were checked regularly and were very stable during the observations.

Our targets were chosen from among the PMS binaries in Taurus-Auriga with the smallest projected separations; we observed 9 binaries with projected separations $a_p \leq 24$ AU. We observed at 800 $\mu$m for greater sensitivity to thermal emission. None of our targets were detected. The observations were calibrated using frequent observations of Mars, CRL 618, and L1551–IRS5. Three-sigma upper limits on the 800-$\mu$m fluxes of the target binaries are given in Table 1.

Interestingly, V773 Tau was not detected in two 990-second observations, one at 800 $\mu$m and one at 1100 $\mu$m, although BSCG reported a flux of 42 $\pm$ 6 mJy at 1300 $\mu$m. V773 Tau has highly variable non-thermal radio emission at 6 cm (Phillips, Lonsdale, & Feigelson 1991), and Skinner, Brown, & Walter (1991) have shown that 1300-$\mu$m emission at the intensity detected by BSCG for V773 Tau could be non-thermal in origin. This is similar to the case of V819 Tau, which was detected by BSCG at 1300 $\mu$m but not detected in subsequent 1100-$\mu$m observations by Skinner et al. (1991). Here we use the upper limits for V773 Tau and V819 Tau as maximum values for their thermal emission, though our conclusions do not change if we use the BSCG fluxes.

---

[3]The JCMT is operated by the Royal Observatory Edinburgh on behalf of the Science and Engineering Research Council of the United Kingdom, the Netherlands Organization for Scientific Research, and the National Research Council of Canada.

## 3. Submillimeter Flux vs. Binary Separation

Our sample is selected from the PMS binaries in the Taurus-Auriga star-forming region, as compiled in Mathieu (1994). We include only binaries with projected angular separations $\leq 10''$, among which contamination from chance superpositions should be negligible (Simon et al. 1992). Our sample is limited to those 42 binaries (out of 60) for which submillimeter fluxes or flux limits are available to us. The flux values and limits used in this analysis are primarily from BSCG and this work, with a few from Skinner et al. (1991), Reipurth & Zinnecker (1993), and Reipurth et al. (1993). There is only one binary in this sample with projected separation less than 1 AU (V826 Tau), so our analyses deal only with binaries having $a_p > 1$ AU. To compare single stars with the PMS binaries, we use the 41 stars in the BSCG sample without detected companions within a projected separation of $10''$ in the surveys of Simon et al. (1992), Ghez et al. (1993), Leinert et al. (1993), Reipurth & Zinnecker (1993), and Richichi et al. (1994). We take projected angular separations from these surveys and Simon (1993, private communication; ZZ Tau) and calculate projected linear separations assuming a distance of 140 pc to the Taurus-Auriga star-forming region (Elias 1978). For the spectroscopic binaries V826 Tau and NTTS 045251+3016 values of $a \sin i$ are taken from Reipurth et al. (1990) and Mathieu (1994), respectively.

For the triple systems UZ Tau, UX Tau, HV Tau, and RW Aur, the entire systems are within the JCMT or IRAM beams; we use the projected separation of the closest pair in each system. For UZ Tau W, we use half of the 1300-$\mu$m flux measured by BSCG based on the equal distribution of flux between the close pair and wider companion in an interferometer map of 2.6-mm continuum emission (Simon & Guilloteau 1992).

In order to compare our new 800-$\mu$m measurements to existing 1300-$\mu$m measurements, we have scaled the 800-$\mu$m fluxes to 1300 $\mu$m by multiplying by $(800/1300)^2$, appropriate for optically-thick emission in the Rayleigh limit. If the emission is partly optically thin, then the flux will decrease more rapidly than $\nu^2$ at longer wavelengths due to the decrease in dust opacity with increasing wavelength. Thus for dust emission our scaling provides upper limits on the 1300-$\mu$m fluxes, preserving the upper-limit nature of our new observations.

Figure 1 shows 1300-$\mu$m flux plotted as a function of projected binary separation $a_p$. A dependence of 1300-$\mu$m flux on projected separation is evident. Nearly all of the detected binaries have $a_p > 100$ AU, and only two binaries with $a_p < 50$ AU are detected (GG Tau and HP Tau). To study the distribution of submillimeter flux with separation, we have used the techniques of survival analysis which are designed for analyzing data that include upper limits.[4] We divided the binaries into two groups based on projected separation and applied various two-sample tests to determine if the fluxes in these samples are drawn from the same distribution. Since 100 AU is often taken as a typical (dust) disk radius (Beckwith & Sargent 1993), we divided the data into binaries with $a_p > 100$ AU and $a_p < 100$ AU. The four two-sample tests give probabilities of 96.6–99.0% that the fluxes in the closer systems are drawn from a different distribution than those in the wider systems. We also applied the tests to the data divided at 50 AU, with the tests giving probabilities 99.1–99.7%. Thus, the distribution of fluxes among the closer binaries is significantly different from that among the wider binaries. In the following statistical analyses we arbitrarily adopt 50 AU as the dividing point between "close" and "wide" binaries, but note that our results do not depend on the specific value between 50 AU and 100 AU.

Simon et al. (1992) suggested that multiple systems in general have lower submillimeter fluxes than single stars. Comparing single stars and the full sample of binaries, we also find a probability of $\geq 99.6\%$ that their distributions of 1300-$\mu$m fluxes are different. However, when the fluxes of the single stars are compared to those of the wide binaries, the two-sample tests show that they are consistent with being drawn from the same distribution. In contrast, the close binaries have a distribution of submillimeter fluxes that is different from that of the single stars with probability $\geq 99.93\%$.

## 4. Interpretation 1: Reduction in Disk Surface Density

What is the origin of the dependence of submillimeter flux on binary separation? We consider two possibilities, in particular reduction of either disk surface density or disk surface area with smaller binary separation. The two need not be exclusive.

The BSCG models for disks around single stars had continuous, power-law distributions of temperature and surface density. For each star/disk the temperature distribution was set by the observed 10–100 $\mu$m spectral energy distribution. The power-law exponent of the surface-density distribution (−1.5) and radius of the disk (100 AU) were taken to be the same for every case. Thus the only free parameter to be set by 1300-$\mu$m flux was the disk surface-density normalization, or equivalently the disk mass. With these constraints, BSCG and Beckwith & Sargent (1993) interpreted the lower submillimeter fluxes of close binaries as indicative of lower disk masses.

---

[4] We used the software package ASURV Rev 1.1 (LaValley, Isobe, & Feigelson 1992) which implements the methods presented in Feigelson & Nelson (1985).

In order to compare our disk mass limits with those found by BSCG in their larger sample, we adopt their disk model, disk temperature parameters (plus disk temperatures derived by us for four binaries), and opacity law which at 800 $\mu$m gives $\kappa_\nu = 0.0375$ cm$^2$ g$^{-1}$. Our $3\sigma$ flux limits at 800 $\mu$m then provide $3\sigma$ upper limits for the disk masses $M_d$, shown in Table 1. Given this model, our 800-$\mu$m flux limits imply disk mass limits of order 0.001 $M_\odot$ for the close binaries. These limits are significantly lower than those derived by BSCG.

The disk masses and mass upper limits of the binaries are plotted in Figure 2 as a function of projected separation. (Five of the binaries plotted in Figure 1 (three close and two wide) and two single stars are excluded from analysis of the mass-separation distribution since there are insufficient infrared data to allow calculation of their disk temperatures and thus disk masses. None of the seven systems are submillimeter detections.) The disk masses show a trend with projected separation similar to that found for the submillimeter fluxes, namely that most of the upper limits found for disk masses among the close binaries are smaller than most of the measured disk masses among the wide binaries. Applying two-sample tests to these data as described above, we find that the distribution of masses among the wide binaries differs from that among the close binaries with a probability of 98.0–99.6%. The distribution of disk masses for the wide binaries is consistent with that of the single stars, while the distributions of disk masses for the close binaries and single stars differ with probability $\geq 99.9\%$.

We note that these specific disk mass estimates are suspect. The assumed continuity in radius of disk material is certainly wrong in the presence of an embedded binary, as such binaries will dynamically clear gaps in the disks. Our calculations of disk masses assuming disks with gaps show that the relatively modest clearing expected from binaries with circular orbits (see next section) does not significantly change the derived masses, but the more extensive clearing expected from binaries with eccentric orbits does have a large effect on the mass derived from a given flux. In addition, disk mass calculations will be inaccurate if binary/disk interactions establish disk temperature and surface-density distributions that differ from those assumed here by analogy to single stars.

## 5. Interpretation 2: Dynamical Clearing of Disks

A binary system will rapidly truncate both circumstellar and circumbinary disks (e.g. Lin & Papaloizou 1993). The resultant disk radii will depend on the details of a given system (e.g., the stellar mass ratio, orbital eccentricity, etc.). However, calculations by Artymowicz & Lubow (1994) suggest that typically the circumstellar disk will have an outer radius of less than half the binary semi-major axis and the circumbinary disk will have an inner radius greater than twice the binary semi-major axis.

Such limits on disk radii decrease submillimeter emission as a consequence of reduced surface area. Indeed, disk truncation leads naturally to a dependence of submillimeter flux on binary separation similar to that seen in Figure 1. Consider a simple picture in which disk formation produces disks with a characteristic outer radius $R$. In wide binaries with separations much greater than $R$, all disks will be circumstellar and little influenced by the presence of companions. These disks and their submillimeter emission would be similar to those around single stars, as observed. In very close systems, only the innermost disk material will be affected by resonant clearing, permitting extensive circumbinary disks with consequent submillimeter emission (e.g. HP Tau, BSCG; GW Ori, Mathieu et al. 1994; AK Sco and V4046 Sgr, Jensen, Mathieu, & Fuller 1994, in preparation). However, for binaries with separations that are a factor of a few less than $R$ a large fraction of the disk area will be cleared, reducing the submillimeter emission. As a specific example, consider a disk with $M_d = 0.01\ M_\odot$. In the BSCG model 75–85% of its 800- and 1300-$\mu$m flux originates between radii of 1 and 50 AU. Thus only if clearing occurs in this part of the disk will the submillimeter flux be reduced substantially.

We have explored this scenario with a simple modification of the BSCG disk model. Specifically, we considered disks having outer radii of 100 AU, a range of temperature distributions, and masses of 0.01 $M_\odot$, typical of the disk masses among the wider binaries. We then removed an annulus of emission to mimic disk clearing by an embedded binary and computed the disk's 1300-$\mu$m flux.[5] We used a range of binary separations and considered two degrees of disk clearing. In one, the clearing extended from 0.5 to 2 times the binary separation, appropriate for a binary with a circular orbit and the minimum clearing expected for any binary system. In the second, the clearing extended from 0.2 to 3 times the binary separation, appropriate for a binary with an eccentricity of 0.4 (Artymowicz & Lubow 1994), roughly the mean eccentricity for PMS and MS binaries with known orbits (Mathieu 1992, Duquennoy & Mayor 1991). (Note that in this simple model only binaries with separations less than a few tens of AU retain substantial circumbinary disks; in reality, dynamical clearing may push material into large circumbinary disks (Pringle 1991).)

We find that clearing from $0.5a$ to $2a$ (the circular-orbit case) reduces 1300-$\mu$m flux by a factor of two at most, insufficient to be consistent with the observed upper limits. However, for clearing expected for eccentric systems ($0.2a$ to $3a$) reductions in flux up to factors of 4–5 are obtained, with the minimum fluxes occurring for binaries in the range 10–50 AU. This is marginally sufficient to reproduce the difference in fluxes between the close binaries and those binaries with separations of 100–500 AU. This result indicates that the close binaries plausibly can have circumstellar or circumbinary disks with surface densities, temperatures, and opacities that are typical for single stars or wide binaries, with the paucity of 1300-$\mu$m flux arising primarily from reduced emitting surface area due to their dynamical truncation.

---

[5] All conclusions stated for 1300-$\mu$m fluxes also apply to 800-$\mu$m fluxes.

This picture alone is unlikely to be a complete explanation of the low fluxes from close binaries. First, dynamical clearing cannot reconcile the flux upper limits of the close binaries with the fluxes of the systems that are brightest in the submillimeter, such as GG Tau and T Tau. In fact, neither GG Tau nor T Tau has an environment as simple as the one envisioned here (e.g. Dutrey, Guilloteau, & Simon 1994, Weintraub et al. 1992). Second, the required gaps are large, implying substantial eccentricities in all PMS binaries. Finally, the flux difference between close and wide binaries is just at the limit of what can be explained by gaps alone. If the flux upper limits for the close binaries are reduced further, another factor must be invoked to explain the observations.

The distribution of disk material resulting from dynamical clearing is certainly more varied and complex than this simple gap model (e.g. Dutrey et al. 1994). Nonetheless, the essential point remains that the present upper limits are consistent with embedded binaries clearing large areas in associated disks while leaving other parts of the disks with large surface densities. Indeed, some circumstellar material is required by the infrared excesses of most binaries at all separations, and the submillimeter data comfortably permit these binaries to have tidally truncated circumstellar disks with surface densities similar to disks around single stars.

## 6. Summary and Discussion

We have placed sensitive upper limits on 800-$\mu$m continuum emission from pre-main sequence binaries in Taurus-Auriga with projected separations less than 25 AU. Combining these limits with previous 1300-$\mu$m observations, we find that submillimeter continuum emission is significantly lower from binaries with projected separations $1 < a_p < 50$ AU than from wider binaries, whose fluxes are comparable to those of single stars. While neither limit on the range of projected separation is well constrained, the binaries showing reduced submillimeter emission have separations less than the radii typically found for disks around single stars. Thus the lack of submillimeter emission must reflect on binary-disk interactions, either at formation or in later evolution.

– 8 –

Lower disk masses among the closer binaries could be the result of consumption of disk material during formation of a binary companion, for example from a gravitational instability in a circumstellar disk (Adams et al. 1989, BSCG). Similarly, formation by disk-enhanced capture (Clarke & Pringle 1991, 1993) might disrupt the disks. Since these mechanisms presumably only form binaries with separations less than typical disk radii, they could account for lower disk masses associated with the closer binaries.

Alternatively, later evolution of binary-disk systems could establish the observed flux-separation relation. One possibility is dynamical clearing of disks, as discussed here. In addition, a companion can increase circumstellar disk accretion rates, possibly lowering the disk mass on timescales comparable to PMS binary ages (Ostriker, Shu, & Adams 1992). The strength of this effect decreases with increasing separation, so disks in wide binaries would be unaffected.

We thank Mike Simon for communicating data in advance of publication. This research has made use of the Simbad database, operated at CDS, Strasbourg, France. ELNJ gratefully acknowledges the support of the National Space Grant College and Fellowship Program, and the Wisconsin Space Grant Consortium. RDM appreciates funding from the Presidential Young Investigator program, a Guggenheim Fellowship, the Morrison Fund of Lick Observatory, and the Wisconsin Alumni Research Fund. GAF acknowledges the support of an NRAO Jansky Fellowship.

## REFERENCES

Abt, H. A. 1983, ARA&A, 21, 343

Adams, F. C., Ruden, S. P., & Shu, F. H. 1989, ApJ, 347, 959

Artymowicz, P., & Lubow, S. H. 1994, ApJ, 421, 651

Beckwith, S. V. W., Sargent, A. I., Chini, R. S., & Güsten, R. 1990, AJ, 99, 924 (BSCG)

Beckwith, S. V. W., & Sargent, A. I. 1993, in *Protostars and Planets III*, ed. E. H. Levy & J. I. Lunine (Tucson:University of Arizona Press), 521

Bonnell, I., Arcoragi, J., Martel, H., & Bastien, P. 1992, ApJ, 400, 579

Clarke, C. J., & Pringle, J. E. 1991, MNRAS, 249, 584

Clarke, C. J., & Pringle, J. E. 1993, MNRAS, 261, 190

Duquennoy, A., & Mayor, M. 1991, A&A, 248, 485

Dutrey, A., Guilloteau, S., & Simon, M., 1994, A&A, in press

Elias, J. H. 1978, ApJ, 224, 857

Feigelson, E. D., & Nelson, P. I. 1985, ApJ, 293, 192

Ghez, A. M., Neugebauer, G., & Matthews, K. 1993, AJ, 106, 2005

Herbig, G. H., & Bell, K. R. 1988, Lick Observatory Bulletin, No. 1111

Jensen, E. L. N., Mathieu, R. D., & Fuller, G. A. 1994, in preparation

Larson, R. B. 1990, in *Physical Processes in Fragmentation & Star Formation*, ed. R. Cappuzo-Dolcetta, C. Chiosi, & A. DiFazio (Dordrecht:Kluwer), 389

LaValley, M., Isobe, T., & Feigelson, E. D. 1992, BAAS, 24, 839

Leinert, Ch., Zinnecker, H., Weitzel, N., Christou, J., Ridgway, S. T., Jameson, R., Haas, M., & Lenzen, R. 1993, A&A, 278, 129

Lin, D. N. C., & Papaloizou, J. 1993, in *Protostars and Planets III*, ed. E. H. Levy & J. I. Lunine (Tucson:University of Arizona Press), 749

Mathieu, R. D. 1992, in *Binaries as Tracers of Stellar Formation*, ed. A. Duquennoy & M. Mayor (Cambridge:Cambridge Univ. Press), 155

Mathieu, R. D. 1994, ARA&A, in press

Mathieu, R. D., Adams, F. C., Fuller, G.A., Jensen, E. L. N., Koerner, D. W., & Sargent, A. I. 1994, AJ, submitted

Ostriker, E. C., Shu, F. H., & Adams, F. C. 1992, ApJ, 399, 192

Phillips, R. B., Lonsdale, C. J., & Feigelson, E. D. 1991, ApJ, 382, 261

Pringle, J. 1991, MNRAS, 248, 754

Reipurth, B., Lindgren, H., Nordstrom, B., & Mayor, M. 1990, A&A, 235, 197

Reipurth, B., & Zinnecker, H. 1993, A&A, 278, 81

Reipurth, B., Chini, R., Krügel, E., Kreysa, E., & Sievers, S. 1993, A&A, 273, 221

Richichi, A., Leinert, Ch., Jameson, R., & Zinnecker, H. 1994, A&A, in press

Simon, M. 1993, private communication

Simon, M., Chen, W. P., Howell, R. R., Benson, J. A., & Slowik, D. 1992, ApJ, 384, 212

Simon, M., & Guilloteau, S. 1992, ApJ, 397, L47

Skinner, S. L., Brown, A., & Walter, F. M. 1991, AJ, 102, 1742

Weintraub, D. A., Kastner, J. H., Zuckerman, B., & Gatley, I. 1992, ApJ, 391, 784

---

TABLE 1
Flux limits and disk mass limits

| HBC[a] | Name | Proj. separation (AU) | $F_{\nu, 800\,\mu m}$ (mJy) | $M_d$ ($M_\odot$) |
|---|---|---|---|---|
| | | | \multicolumn{2}{c}{$3\sigma$ limits} | |
| | HQ Tau | 1.26[b] | < 27 | < 0.001 |
| | GN Tau | 5.74[b] | < 31.5 | < 0.001 |
| 36 | DF Tau | 12.3 | < 19 | < 0.001 |
| 39 | DI Tau | 16.8 | < 25.5 | < 0.001 |
| 46 | ZZ Tau | 4.06[b] | < 100.5 | ··· [c] |
| 367 | V773 Tau | 16.0 | < 29[d] | < 0.001 |
| 369 | FO Tau | 23.2 | < 36 | < 0.002 |
| 400 | V826 Tau | 0.06[e] | < 36 | < 0.007 |
| 427 | NTTS 045251+3016 | 4.0[e] | < 42 | ··· [f] |

[a] Number from the catalog of Herbig & Bell (1988).

[b] Separations projected along the direction of lunar occultation.

[c] Flux limit high due to poor sky conditions; $F_{\nu, 1300\,\mu m} < 15$ mJy, $M_d < 0.014\ M_\odot$ (BSCG) used in analysis.

[d] Also $F_{\nu, 1100\,\mu m} < 36$ mJy ($3\sigma$).

[e] Projected semi-major axis $a \sin i$ from spectroscopic orbit.

[f] Insufficient infrared data to calculate $M_d$.

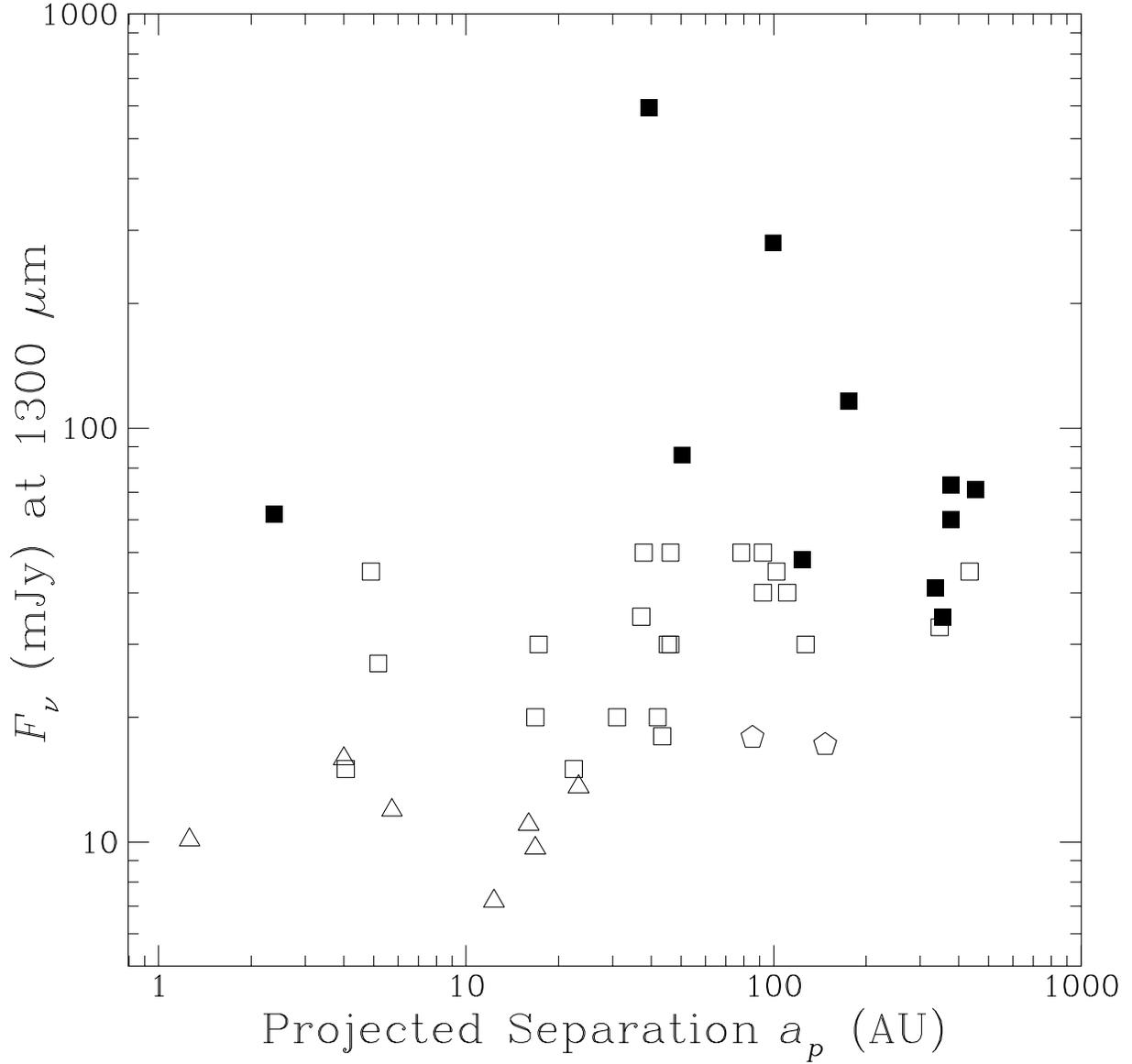

Fig. 1.— Plot of 1300-$\mu$m flux vs. projected binary separation $a_p$ for young binaries in Taurus-Auriga. Filled squares give fluxes for detected binaries; open symbols are $3\sigma$ upper limits for systems not detected. Squares show fluxes from 1300-$\mu$m observations, triangles are 800-$\mu$m fluxes scaled to 1300-$\mu$m, and pentagons are 1100-$\mu$m fluxes scaled to 1300-$\mu$m. Note the lack of detections among the closer binaries.

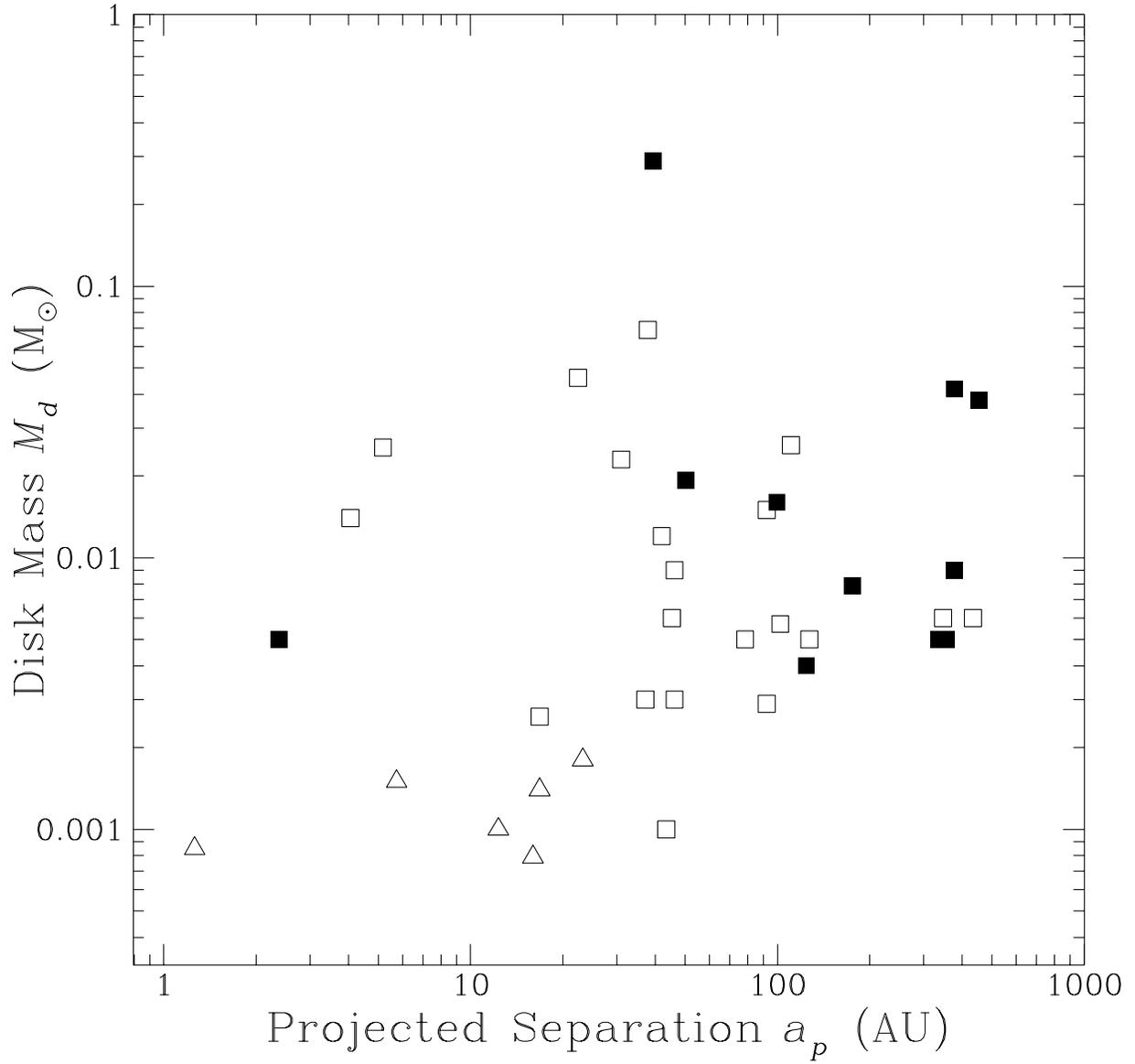

Fig. 2.— Plot of disk mass $M_d$ vs. projected binary separation $a_p$ for young binaries in Taurus-Auriga. Symbols are as in Figure 1. Note the lack of massive disks in systems with separations < 50 AU.